\documentclass[prb,preprint]{revtex4} 

\usepackage{amsmath}
\usepackage{graphicx}

\begin{document}

\title{Periodic orbits, localization in normal mode space,
and the Fermi-Pasta-Ulam problem}

\author{S. Flach}
\email{flach@mpipks-dresden.mpg.de}
\affiliation{Max-Planck-Institut f\"ur Physik komplexer Systeme
N\"othnitzer Str. 38, D-01187 Dresden, Germany}
\author{M. V. Ivanchenko}
\affiliation{Department of Applied Mathematics, University
of Leeds, Leeds, LS2 9JT, United Kingdom}
\author{O. I. Kanakov}
\author{K. G. Mishagin}
\affiliation{Department of Radiophysics, Nizhny Novgorod University
Gagarin Avenue 23, 603950 Nizhny Novgorod, Russia}


\begin{abstract}
The Fermi-Pasta-Ulam problem was one of the first
computational experiments. It has stirred the physics community
since, and resisted a simple solution for half a century.
The combination of straightforward simulations, efficient
computational schemes for finding periodic orbits, and analytical
estimates allows us to achieve significant progress.
Recent results on $q$-breathers, which are time-periodic solutions that
are localized in the space of normal modes of a lattice and
maximize the energy at a certain mode number, are
discussed, together with their relation to
the Fermi-Pasta-Ulam problem. The localization properties
of a $q$-breather are characterized by 
intensive parameters, that is, energy densities and wave numbers.
By using scaling arguments, $q$-breather solutions are 
constructed in systems of arbitrarily large size.
Frequency resonances in certain
regions of wave number space lead to the complete delocalization
of $q$-breathers. The relation of these features to
the Fermi-Pasta-Ulam problem
are discussed.
\end{abstract}

\maketitle

\section{Introduction}
\label{sec1}
The characterization of dynamical excitations of a system is one of the
fundamental tasks in condensed matter physics. For
quantum systems the problem is obtaining the correct description
of the many-particle wave functions.
For many-particle nonlinear classical systems
the main difficulty is relating the excitations to families of trajectories in phase
space.

Because a generic classical Hamiltonian system is non-integrable, it
will evolve chaotically on large observation times for almost any
initial condition. Hamiltonian chaos describes the decay of
excitations on large time scales. The excitations are
typically characterized by nonchaotic trajectories. 
Kolmogorov, Arnold, and Moser (KAM) showed that regular dynamics
on invariant $N$-dimensional tori persists in many non-integrable
systems with a finite number of degrees of freedom $N$.\cite{via89} Especially important are the simplest realizations
of low-dimensional invariant structures -- periodic orbits. 
Periodic orbits
exist even in strongly chaotic systems and can densely fill the
chaotic phase space volume despite having zero measure (just like
the rational numbers fill the space of real numbers). 
Periodic orbits are the most natural class of regular trajectories which
can describe excitations. An example of such
excitations are discrete breathers. They are periodic solutions of
nonlinear lattice systems localized in the direct space that
persist in infinite lattices.\cite{p1,fw98pr,fg05chaos} 

Periodic orbits can be computed numerically with high accuracy for
large systems. Therefore computational approaches can often help
in characterizing dynamical excitations, especially when combined
with analytical arguments. In the following we will show
how such methods allow us to successfully study one of the most 
famous problems in computational and statistical physics -- the Fermi-Pasta-Ulam problem.

\section{The Fermi-Pasta-Ulam problem}
\label{sec2}

In 1953 Enrico Fermi, John Pasta, and Stanislav Ulam studied
the problem of energy redistribution between the linear normal
modes of an anharmonic atomic chain and published the results as an
internal report of the Los Alamos National Laboratory. \cite{p2}
They used one of the first available computers with modern architecture,
and performed what can be called one of the first 
computational experiments.

The equations of motion describe $N$ particles with a nonlinear
interaction between nearest neighbors, with either
quadratic (the $\alpha$-FPU model)
\begin{equation}
\label{eq1}
\ddot{x}_n=(x_{n+1}-2x_n+x_{n-1})
+\alpha\big[(x_{n+1}-x_n)^2-(x_n-x_{n-1})^2 \big],
\end{equation}
or cubic terms (the $\beta$-FPU model)
\begin{equation}
\label{eq2}
\ddot{x}_n=(x_{n+1}-2x_n+x_{n-1})
+\beta \big[(x_{n+1}-x_n)^3-(x_n-x_{n-1})^3 \big].
\end{equation}
The coordinate $x_n(t)$ corresponds 
to the displacement of the
$n$th particle from its equilibrium position.
The particle mass and harmonic spring constant
are assumed to be one. 

\begin{figure}
{
\resizebox*{0.45\columnwidth}{!}{\includegraphics{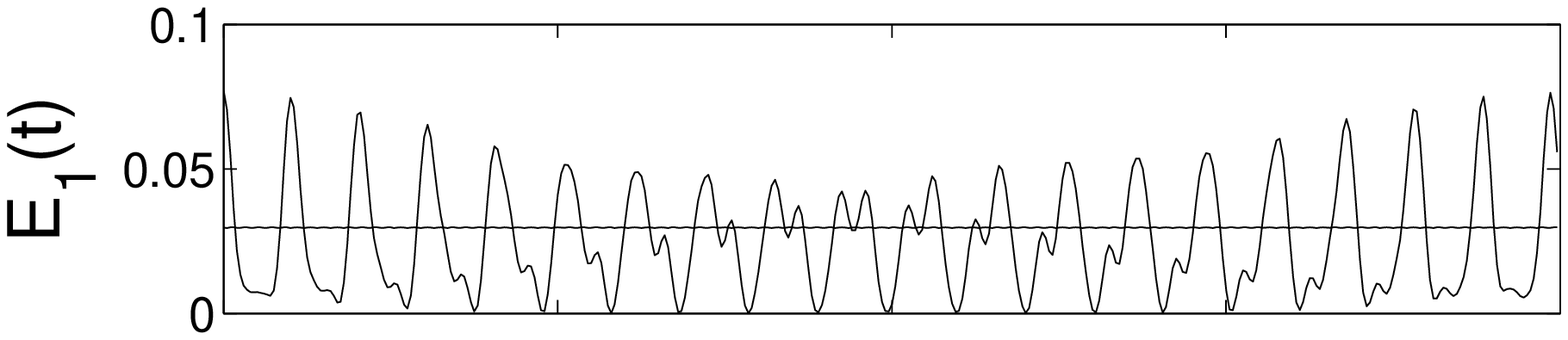}}\hfill
\resizebox*{0.45\columnwidth}{!}{\includegraphics{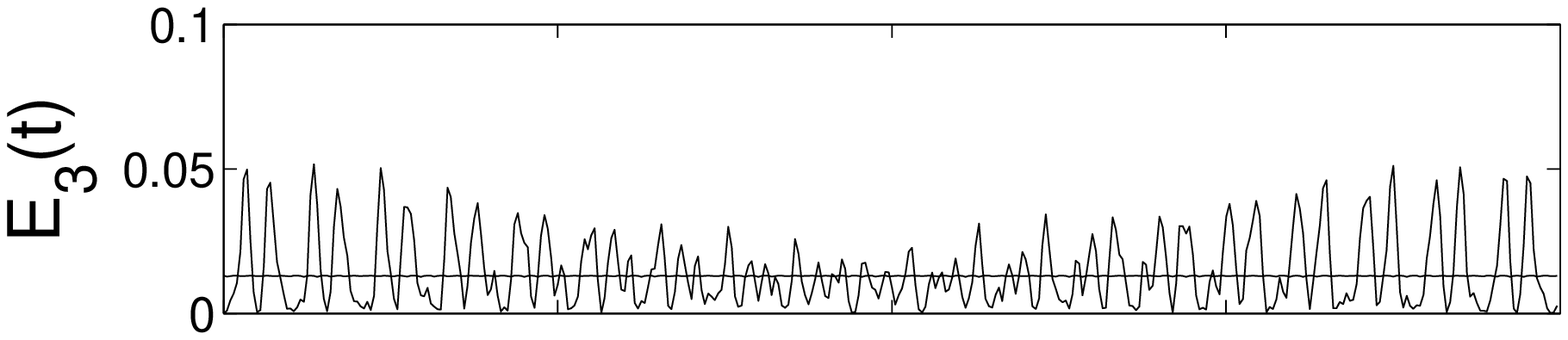}}\\
\resizebox*{0.45\columnwidth}{!}{\includegraphics{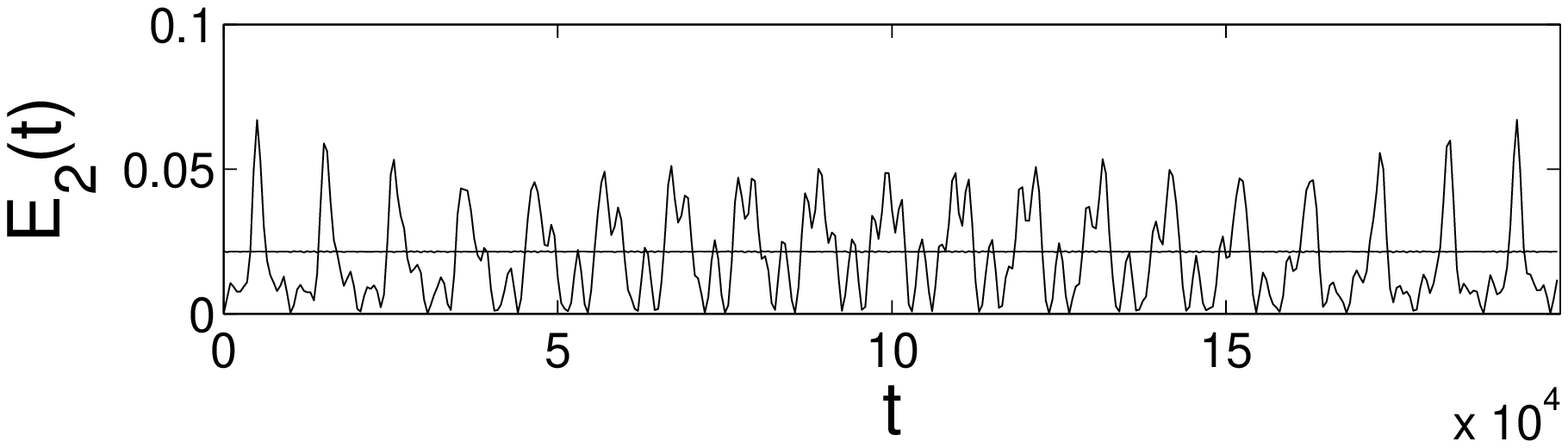}}\hfill
\resizebox*{0.45\columnwidth}{!}{\includegraphics{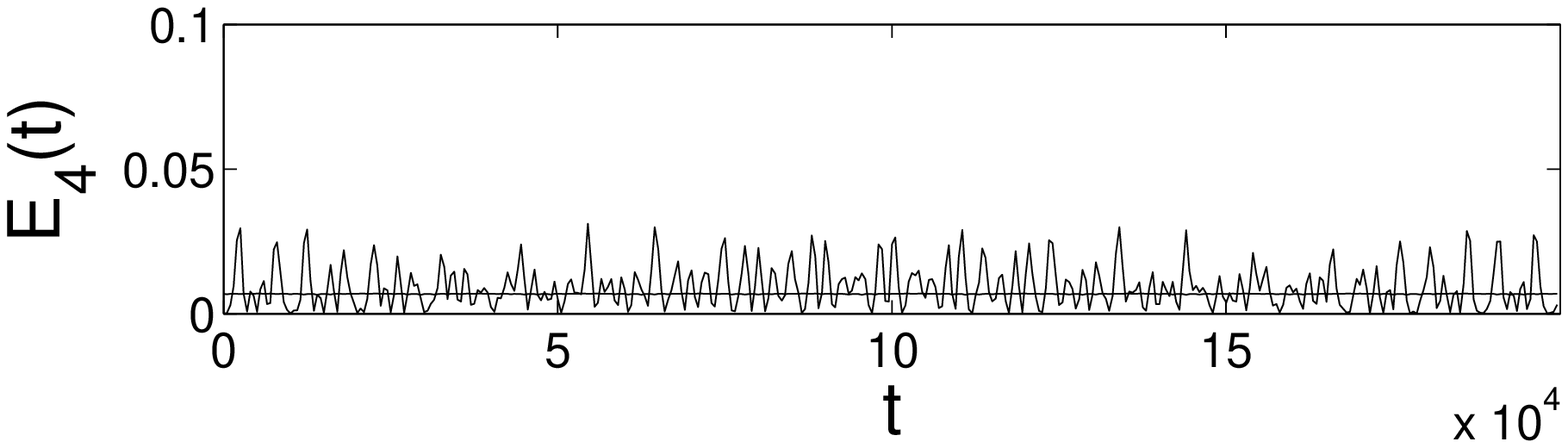}}}
\caption{The evolution of the energy of modes $q=1,\,2,\,3,\,4$ for
the FPU trajectory (oscillating curves) and the exact periodic orbit of
the $q$-breather (horizontal lines). The parameters are $\alpha=0.25$,
total energy $E=0.077$, and 
$N=32$.\cite{p3}} \label{Fig1}
\end{figure}

Fixed boundary conditions $x_0=x_{N+1}=0$ are used. The
canonical transformation 
\begin{equation}
x_n(t)=\sqrt{\frac{2}{N+1}}\sum\limits_{q=1}^N
Q_q(t)\sin{\Big(\frac{\pi q n}{N+1} \Big)}
\end{equation}
diagonalizes and solves the linear problem $\alpha=\beta=0$ 
using the normal mode coordinates
$Q_q(t)$. The mode number $q=1, \ldots,N$ relates each of these modes to their corresponding
normal mode frequency $\omega_q=2\sin{(\pi q/2(N+1))}$. The
equations of motion in normal mode space become
\begin{equation}
\label{eq3}
\ddot{Q}_q+\omega_q^2 Q_q=-\frac{\alpha}{\sqrt{2(N+1)}}\sum\limits_{l,m=1}^N
\omega_q\omega_l\omega_m B_{q,l,m} Q_l Q_m
\end{equation}
for the $\alpha$-FPU chain in Eq.~(\ref{eq1}) and
\begin{equation}
\label{eq4}
\ddot{Q}_q+\omega_q^2 Q_q=-\frac{\beta}{2(N+1)}\sum\limits_{l,m,n=1}^N
\omega_q\omega_l\omega_m\omega_n C_{q,l,m,n} Q_lQ_mQ_n
\end{equation}
for the $\beta$-FPU chain in Eq.~(\ref{eq2}). The interaction coefficients
\begin{equation}
\label{eq5}
B_{q,l,m}=\sum\limits_{\pm}(\delta_{q\pm l \pm m,0}-
\delta_{q\pm l \pm m,2(N+1)}),
\end{equation}
\begin{equation}
\label{eq6}
C_{q,l,m,n}=\sum\limits_{\pm}
(\delta_{q\pm l \pm m \pm n,0}-\delta_{q\pm l \pm m \pm n,2(N+1)}
-\delta_{q\pm l \pm m \pm n,-2(N+1)})
\end{equation}
give the coupling between the modes. The coupling is nonlinear, selective,
and does not have a 
characteristic interaction range (or distance) in normal mode space. 
Thus, even very distant modes in normal mode space are coupled.
In the
absence of this interaction there exist $N$ integrals of motion
which correspond to the energies of the normal modes
$E_q=(\dot{Q}_q^2+\omega_q^2 Q_q^2)/2$.

\begin{figure}
{\includegraphics[angle=-90,width=0.8\columnwidth]{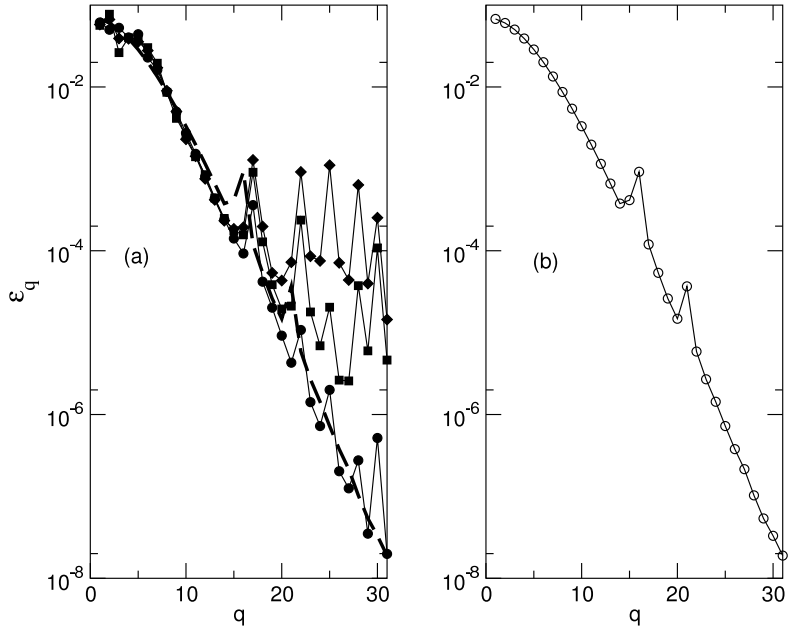}}
\caption{(a) Distributions of the mode energy densities for the
FPU trajectory with $q_0=1$, $N=31$, $\alpha=0.33$, and $E=0.32$. 
Circles: $t=10^4$, squares: $t=10^5$, diamonds: $t=10^6$. The dashed line
is the $q$-breather from (b) for comparison. (b) Distributions of the mode
energy densities for the $q$-breather with the same parameters as in (a).
\cite{p4}} \label{Fig2}
\end{figure}

Fermi, Pasta, and Ulam considered the evolution of the nonlinear chain with the 
single mode $q_0=1$ with the lowest frequency initially excited.\cite{p2} 
They used standard numerical schemes to approximate
the differential equations and monitored the
state of the system for times which are orders of magnitude larger 
than the largest periods of oscillation of the normal modes. 
It was expected that after some 
time other
modes would become excited as well, and the energy would eventually be
equally distributed over the entire spectrum. That was the
way they 
anticipated a transition to thermal equilibrium. 
To their big surprise they observed
the opposite result (see Fig.~\ref{Fig1}). The energy stayed localized
in several low frequency modes for times that are orders of magnitude
larger than the typical oscillation periods of these normal modes.
Recurrences of
almost all the energy into the initially excited mode were observed as well.
Later computational studies
revealed that there exist energy thresholds and system size
thresholds above which equipartition was observed on relatively
short time scales.\cite{Sh,p5} Many questions arise.
Why is the energy localized in a few modes below the thresholds? Why
does the energy spread quickly into the entire spectrum above the thresholds?
How do the threshold values scale with the parameters of the system?

Figure~\ref{Fig2} shows the mode energy density distribution
$\varepsilon_q = E_q/N$ for an initial energy that is larger than the 
one considered in Ref.~\onlinecite{p2}, but yet well below the above mentioned
thresholds. The distributions are plotted on a
logarithmic scale and for times $t=10^4,\,10^5$, and $10^6$. Because the
mode energies are time-dependent, the densities were averaged over
a time interval of $\Delta t = 10^4$. We observe
exponential localization of the energy distribution in normal mode space,
which was first noted by Galgani and Scotti.\cite{gs72}
Such a distribution is formed on a very short time scale
$\tau_1$,and 
possesses a core --- the few modes which are strongly excited ---
and a tail of exponentially weakly excited modes.
Also shown in Fig.~\ref{Fig2} is a slow resonant pumping of energy
from the core of the distribution into its tails for later times. 
This process ultimately brings the system to
equipartition, but on the much larger time scale $\tau_2 \gg \tau_1$. 
Fermi, Pasta, and Ulam\cite{p2} performed their numerical studies at lower energies, and therefore
$\tau_2$ was even larger in their case. 
If the initial energy is further increased, a threshold value may be reached
for which the
time scale of the core-tail pumping becomes comparable with the
transition time to the exponential distribution $\tau_2
\approx \tau_1$.\cite{Sh,p5} Below these threshold values the time scales 
differ by orders of magnitudes, and the evolution on times $\tau_1 \ll t \ll \tau_2$
corresponds to the observations of Ref.~\onlinecite{p2}. Fermi, Pasta, and Ulam viewed their observations
as a problem of
(perhaps complete) the absence of equipartition (that is, ergodicity) -- a problem now known as the FPU problem.
From a contemporary perspective, the FPU problem can be reformulated as the
following set of questions:
\begin{itemize}
\item
Why are there at least two qualitatively different time scales $\tau_1$ and $\tau_2$?
\item
Are there low-dimensional invariant manifolds in phase space
which approximate the dynamics of the FPU trajectories for times
less than $\tau_2$?
\item
Can we estimate the localization length of the exponentially localized energy
distribution in normal mode space?
\item
Is the localization length 
related to the time scales?
\item
Does the FPU observation hold in the
limit of infinite systems, as well as in two- and three-dimensional
systems?
\end{itemize}

There is a long history of studies of the FPU problem, \cite{njzmdk65,IC,Ford92}
which, even if not directly answering all of these questions, 
defined new fields of research in nonlinear dynamics, chaos theory, and soliton theory.

\section{Periodic orbits -- $q$-breathers}
\label{sec3}

From the results of Sec.~\ref{sec2} we may conjecture that
the exponentially localized state observed on the FPU trajectory
is a long-lived excitation, which is destroyed after time $\tau_2$.
Therefore (see discussion in Sec.~\ref{sec1}) we anticipate that this excitation 
is close to an exact periodic orbit for times $t \ll \tau_2$. 
Then the observed
dynamics will be almost regular. Such a periodic orbit
will correspond to a slightly deformed normal mode. There exists a rigorous method of
constructing such a periodic orbit, starting from the linear limit
$\alpha=\beta=0$.\cite{p3,p6} In this limit the periodic orbit is given by exciting
a single mode with mode number $q_0$
with an energy $E$. With these conditions 
we isolate
a unique periodic orbit in the
multi-dimensional phase space of the system.

The linear spectrum
of the FPU chain obeys the non-resonance condition\cite{p7}
\begin{equation}
\label{eq7} \omega_q \neq n \omega_{q_0}
\end{equation}
for all integer $n$ and $q \neq q_0$. This condition allows us to
apply a theorem due to Lyapunov,\cite{p8} which states that
the original periodic orbit can be continued into the domain of nonzero nonlinearity
for fixed energy.\cite{p3} 
Continuation means that close to the periodic orbit of the linear system we
will find a new periodic orbit in the phase space of the nonlinear system with
identical energy $E$. This new periodic orbit, being a closed loop in phase space,
can be viewed as a continuous deformation of the periodic orbit of the linear system,
as the nonlinearity is increased. 

The numerical methods of constructing $q$-breathers
\cite{p3,p6} use the fact that a periodic orbit corresponds to a fixed point
in the generalized Poincar\'e
maps (sections) in phase space. 
Let us explain one of the possible realizations of such a numerical scheme in more detail. 

We consider initial conditions for which all normal mode velocities $\dot{Q}_q(t=0)=0$.
Then the kinetic energy at $t=0$ is exactly zero. 
Note that the kinetic energy is given by the one-half of the sum over all squared mode velocities, and the total energy is conserved
during the integration.
We construct a $N$-dimensional vector 
where each component
is the amplitude of a given mode. This vector represents the configuration space of the system. 
Then we integrate the equations of motion numerically, until the velocity $\dot{Q}_{q_0}$
crosses zero two times (because an oscillator needs two turning points to return
to its initial state).
At that moment we again measure all mode amplitudes, and obtain
a second (final) vector. If the initial 
and final vectors coincide, the potential
energies coincide as well. Therefore the kinetic energy at the final time will be zero,
and our initial condition is located on a periodic orbit. The period of the orbit
can be determined by computing the time we needed to reach the final vector.

What if the final vector does not coincide with the initial one? Then we compute the
difference vector between the final and initial vectors. 
It will also have $N$ components. Our task is to slightly
vary the initial vector components such that all the components of the difference vector
shrink to zero. In other words, we are trying to zero $N$ functions (the components
of the difference vector) by varying $N$ variables (the components of the initial vector).
Generalized Newton-Raphson methods will do the job if the initial vector is suitably
close to a desired solution. Because the final vector is the outcome of a map in configurational
space, we can also say that we are looking for a fixed point of that map.
If we want to find a fixed point for a given total energy (or some other constraint)
we have to limit the variations of the initial vector accordingly. 
For practical reasons it is convenient to first fix the amplitude of mode
$Q_{q_0}$ to a given value $a$ to find the corresponding periodic orbit,
calculate its energy, and then to vary $a$ such that the desired energy value is realized. 

Once a periodic orbit is obtained, we also compute its stability properties.
A periodic orbit is said to be (marginally) stable if a small perturbation
of the orbit in phase space does not grow in time; otherwise it is unstable.
The initial smallness of the perturbation allows us to perform a linearization of
the phase space flow around the periodic orbit (see Ref.~\onlinecite{via89} for details).
In practice it amounts to the following procedure. Take a reference point on the periodic orbit.
Consider $2N$ small perturbations in the $2N$ different directions in phase space.
Integrate each of these perturbations over one period of the orbit, and obtain
a new (still quite small) perturbation. The components of these final perturbations
define a Floquet matrix, which maps any initial perturbation into a final one.
The matrix is symplectic.\cite{sf04}
In a nutshell, if $\lambda$ is an eigenvalue of a symplectic
matrix, so are $1/\lambda, \lambda^*$ and $1/\lambda^*$. 
Compute the eigenvalues of that matrix. Any eigenvalue
which is located on the unit circle in the complex plane corresponds to a perturbation
that does not grow in time. Eigenvalues outside the unit circle 
correspond to perturbations which grow exponentially fast in time. Note that each eigenvalue
outside the unit circle has a symmetry-related partner inside the unit circle due to
the symplectic properties of the Floquet matrix.
The interested reader may consult Ref.~\onlinecite{sf04}, where details of the ways to compute
and analyze periodic orbits are discussed.

\subsection{Results for the $\alpha$-FPU model, low frequencies}
\label{sec3.1}

The dynamics of the continued
periodic orbit incorporates many normal modes, because nonlinearity
induces mode-mode interactions. 
The numerical procedure described in Sec.~\ref{sec2} was used and new periodic
orbits were found.
The amplitude
and energy distributions of the normal modes in such a periodic orbit turn
out to be exponentially localized
(see Figs.~\ref{Fig1} and \ref{Fig2}(b)). We call these solutions $q$-breathers;
they are periodic in time and localized in normal mode
space. Note that the shape of the energy distribution of $q$-breathers
is remarkably close to the one of the FPU-trajectory for $t < \tau_2$ and
identical parameters (energy $E$ and seed mode number $q_0$).

Any numerical evaluation should, if possible, be accompanied by 
analytical considerations. For $q$-breathers that is possible
with the help of standard perturbation approaches.
For the case of
low-frequency modes the energy distribution in
the $q$-breather has been estimated to be\cite{p6}
\begin{equation}
\label{eq8} E_{q=n q_0}=\gamma^{2n-2}n^2 E_{q_0}, \quad
\gamma=\frac{\alpha \sqrt{E_{q_0}}(N+1)^{3/2}}{\pi^2
q_0^2},
\end{equation}

\begin{figure}
\includegraphics[width=0.6\columnwidth]{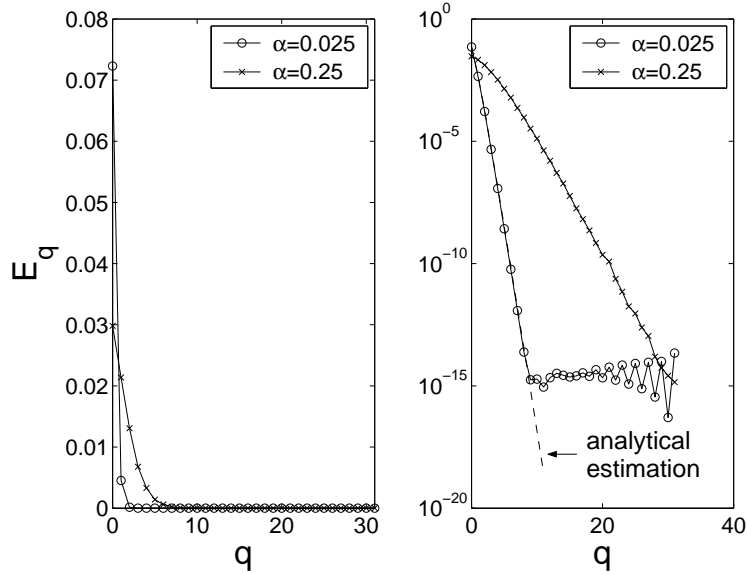}
\caption{Stable $q$-breathers for $\alpha=0.025$ and $\alpha=0.25$,
$E=0.077$, $N=32$, and $q_0=1$. The dashed line is the result of
Eq.~(\ref{eq8}).\cite{p6}} \label{Fig3}
\end{figure}

The $q$-breathers are localized in normal mode space if mode energies shrink with
increasing distance from $q_0$.
From Eq.~\eqref{eq8} it follows that $q$-breathers are exponentially localized if
$\gamma < 1$: $E_q \sim e^{-q/\xi}$, where $\xi$ is the localization length.
For 
$\gamma=1$
Eq.~(\ref{eq8}) predicts the parameter values
for which a $q$-breather delocalizes. If this condition is reached upon
variation of a parameter, we say that a delocalization threshold has been reached.
If we vary a second parameter, we can keep $\gamma$ unchanged. In this way we can
vary all parameters, keep $\gamma$ constant, and obtain scaling relations
between all the parameters. 
The inverse localization length $\xi^{-1}$ is
proportional to $q_0^{-1} \ln \gamma$. When $\gamma = 1$ the $q$-breather
delocalizes, which implies $\tau_2 \approx \tau_1$. 
These are, among many others, predictions which can be
tested by comparing to $q$-breather solutions and FPU-trajectories, which are obtained using the
previously described computational methods.

The numerical
results and the estimate in Eq.~(\ref{eq8}) are shown in
Fig.~\ref{Fig3}. The $q$-breather solutions are linearly stable.\cite{p3,p6} 
The agreement between the computations and theory is very good, especially for strongly
localized energy profiles.

Note that in real space the energy distributions are delocalized
both for the FPU and for the $q$-breather trajectories, and do not show 
remarkable
features.\cite{p3,p6}

\subsection{Results for $\beta$-FPU, low frequencies}
\label{sec3.2}

\begin{figure}
\resizebox*{0.5\columnwidth}{!}{\includegraphics{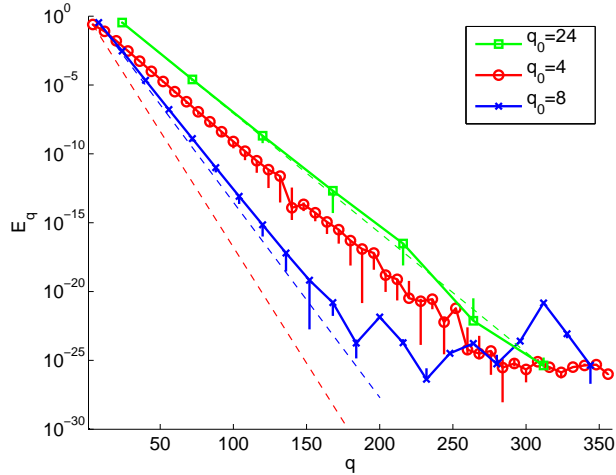}}
\caption{(color online) The profiles of the $q$-breather solutions for $N=359$,
$\beta=1$, $E=0.346$, and $q_0=4$, 8, and 24.
The symbols denote the mode energies at the time when the
coordinates $x_n(t)=0$. The vertical bars denote the
range of mode energy values realized during one period of the
$q$-breather. The dashed lines are the result if Eq.~(\ref{eq9}) with
$q_0=24$, 8, and 4 from top to bottom.\cite{p11}}
\label{Fig4}
\end{figure}

If we apply the perturbation methods we have mentioned to the $\beta$-FPU problem, 
we arrive at the result for the energy distribution in
the $q$-breather:\cite{p6}
\begin{equation}
\label{eq9}
E_{q=(2n+1)q_0}=\lambda^{2n}E_{q_0},\quad
\lambda=\frac{3\beta E_{q_0}(N+1)}{8\pi^2 q_0^2}.
\end{equation}
The necessary condition for the localization of the $q$-breather is 
$\lambda < 1$. The inverse localization length $\xi^{-1}$ is
proportional to $(2 q_0)^{-1} \ln \lambda$. 
If $\lambda = 1$, the $q$-breather
delocalizes, and thus we expect that $\tau_2 \approx \tau_1$. 
The numerical and analytical results (\ref{eq9}) are presented in Fig.~\ref{Fig4}. 
For $q_0=4$ the asymptotic result for $\lambda$ fails, although
the exponential profile of the mode energies still holds.

The $q$-breathers become unstable if $6\beta E(N+1)/\pi^2 > 1$,\cite{p6} but
as long as they remain strongly localized, the FPU trajectories do
not delocalize quickly.\cite{p9} 
Remarkably the FPU trajectory shows a change from
regular to weakly chaotic dynamics in its core when increasing the initial energy $E$
above a certain value.\cite{p9}
This value corresponds to the 
condition for changing the stability of the $q$-breather.
It is well known that for nonintegrable Hamiltonian systems, chaotic trajectories appear
close to an unstable periodic orbit.\cite{ll92}

Why does the instability of the $q$-breather, and the chaotic dynamics close to it,
not destroy the exponential localization of the mode energy profile?
The answer might be that a $q$-breather which is well localized in normal mode space
does not allow for a resonant interaction between core and tail modes,
because if such a resonance was present, it
would lead to a delocalization of the periodic orbit itself. 
Instabilities in the core of the distribution, that is, 
resonances between modes inside the core, are what is left.
They can lead to a chaotic dynamics 
inside the core, 
but not to a delocalization, due to the absence of
resonances between the core and the tail modes. 

\subsection{Results for high frequencies}
\label{sec3.3}

\begin{figure}
{\includegraphics[angle=-90,width=0.8\columnwidth]{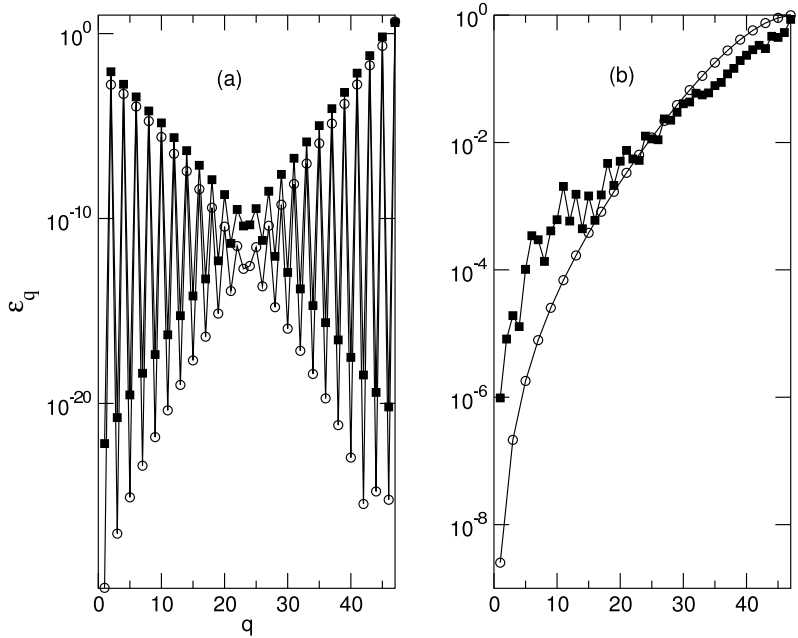}}
\caption{Energy density distribution for the FPU trajectory
(squares, $t=10^5$) and $q$-breather (circles) for $N=47$, $E=4.7$, and
$q_0=47$. (a) $\alpha=0.25$ and (b) $\beta=0.25$.\cite{p4}}
\label{Fig5}
\end{figure}

Localization is observed also for high frequency seed modes.
In Fig.~\ref{Fig5} the energy distributions for the FPU and $q$-breather
trajectories are shown.\cite{p4} The only difference from low frequencies is that 
for the $\alpha$-FPU model high and low
frequency modes become excited in pairs, in agreement with the predictions of
perturbation theory.\cite{p4}

\section{Generalization to two and three dimensions}
\label{sec4}

\begin{figure}
\resizebox*{0.5\columnwidth}{!}{\includegraphics{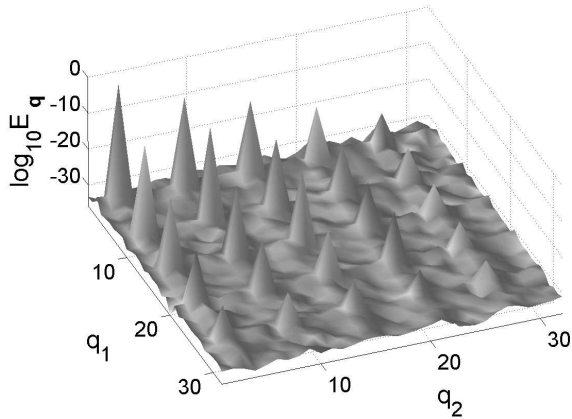}}
\caption{Mode energy distributions for a $q$-breather in a two-dimensional
FPU lattice with $N=32 \times 32$, $E=1.5$, ${\bf q}_0=(3,3)$, and
$\beta=0.5$.\cite{p10}} \label{Fig6}
\end{figure}

The condition for applying Lyapunov's theorem to
the continuation of a $q$-breather is the absence of
resonances (see Eq.~(\ref{eq7})). A finite system has a discrete spectrum and thus
Lyapunov's theorem is applicable. A generalization of the
results in Sec.~\ref{sec3} has been performed by considering
finite lattices with spatial
dimensions $d=2$ and 3. Instead of mode numbers we now have
mode vectors with $d$ integer components. The main properties of the solutions for $q$-breather 
solutions do not change.\cite{p10}
In Fig.~\ref{Fig6} an example of a $q$-breather solution
in $d=2$ is shown.

For the $\beta$-FPU model we can generalize the perturbation theory and predict 
the energy distribution\cite{p10}
\begin{equation}
\label{eq10}
E_{(2n+1){\bf q}_0}=\lambda_{d}^{2n}E_{{\bf q}_0}, \quad
\lambda_{d}=\frac{3\beta
E_{{\bf q}_0}N^{2-d}}{2^{2+d}\pi^2|{\bf q}_0|^2}.
\end{equation}
The numerically computed $q$-breather solutions show quantitative agreement
with these estimates.\cite{p10}

\section{Scaling and transition to macroscopic systems}
\label{sec5}

\begin{figure}
\resizebox*{0.5\columnwidth}{!}{\includegraphics{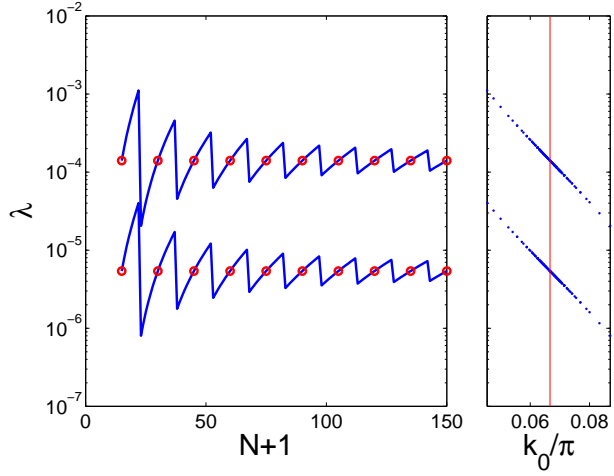}}
\caption{(a) Dependence of $\lambda$ on the size of the chain 
for the $q$-breather with energy density $\varepsilon=4
\times 10^{-4}$ (lower curve) and $2 \times 10^{-3}$ (upper curve)
and $\beta=1$. The open circles 
correspond to $N+1= r (N_0 +1)$ with $N_0=15$,
where $r$ is an integer. (b) The dependence of
$\lambda$ on the wave number $k_0$ for the corresponding data from (a).\cite{p11} } \label{Fig7}
\end{figure}

The perturbation theory results in Eqs.~(\ref{eq8})--(\ref{eq10}) show that
after replacing the
extensive parameters (the energy $E$ and the mode number $q$)
by intensive ones (the energy density $\varepsilon = E/N$ and wave
number $k = \pi q /(N+1)$), the localization length becomes independent
of the system size $N$. Consequently, we expect that $q$-breathers will persist
in the limit of an infinitely large macroscopic system. 
Consider a $q$-breather solution $Q_{q}(t)$ for a finite chain
of size $N$ and another chain of size
$\tilde{N}+1=r( N+1)$, where $r=2,\,3,\,4,\,\ldots$ It follows
that
\begin{equation}
\label{eq11}
\tilde{Q}_{\tilde{q}}(t)=
\begin{cases}
\sqrt{r} Q_q(t), & \tilde{q}=r q
\\
0, & \tilde{q} \neq r q
\end{cases}
\end{equation}
is a solution for the larger chain.\cite{p11}

By increasing $r$ to infinity we will obtain solutions for infinitely
large macroscopic systems. The scaling procedure is easily
generalized to $d=2$ and 3 as well as to free and periodic
boundary conditions.\cite{p11}

Suppose that $q$-breathers exist in an infinitely large chain. Then
Eq.~(\ref{eq9}) with $k$ and $\varepsilon$ instead of $q$ and $E$ gives
\begin{equation}
\label{eq12} \ln \varepsilon_k = \Big(\frac{k}{k_0}-1 \Big) \ln
\sqrt{\lambda} + \ln \varepsilon_{k_0}, \quad
\sqrt{\lambda}=\frac{3\beta}{8} \frac{\varepsilon_{k_0}}{k_0^2}.
\end{equation}
We see that the analytical result for the mode energy profile is independent
of system size if intensive parameters are used.

Consider a finite chain with initial length 
$N_0=15$ and seed wave number
$\bar{k}_0=\pi/15$. We increase the system size and numerically compute
$q$-breather solutions for the seed mode number $q_0$ which is the
closest to $\bar{k}_0(N+1)/\pi$. The
solution is then used to approximate the coefficient $\lambda$ by the
ratio $\varepsilon_{5q_0}/\varepsilon_{3q_0}$ (see Fig.~\ref{Fig7}(a)). 
The results confirm the independence of the approximate
value of
$\lambda$ if $N+1=r(N_0+1)$. For other
values of $N$ we probe $\lambda$ with different seed wave
numbers $k_0$ in the vicinity of $\bar{k}_0$. The continuous behavior of
$\lambda(k_0)$ (see Fig.~\ref{Fig7}(b)),
irrespective of system size $N$, confirms these conclusions for
macroscopic systems. 

If the energy density $\varepsilon$ is fixed to a given value, 
it follows from the geometric
series in Eq.~(\ref{eq9}) that $\varepsilon_{k_0} =
(1-\lambda)\varepsilon$. We use Eq.~(\ref{eq12}) and calculate the slope $S$
of the energy density distribution (on a logarithmic scale):\cite{p11}
\begin{equation}
S= \frac{1}{k_0} \ln \sqrt{\lambda}, \quad \sqrt{\lambda} =
\frac{\sqrt{1+4\nu^4/k_0^4}-1}{2\nu^2/k_0^2}, \quad
\nu^2=\frac{3\beta}{8}\varepsilon. \label{eq13}
\end{equation}
The absolute value of $S$ equals the inverse localization length
$\xi^{-1}=|S|$. The slope $S$ depends on the seed wave number $k_0$ and
on the effective nonlinearity parameter $\nu$, which is determined by the
product of the energy density $\varepsilon$ and the nonlinearity strength
$\beta$. The localization length diverges if $k_0 \rightarrow 0$
and takes a minimum value of $\xi_{\min} \approx \nu/0.7432$ at
$k_0 = k_{\min}\approx 2.577 \nu$.\cite{p11} The
localization length for $k_0=k_{\min}$ also increases for increasing $\nu$. In the limit
$k_0 \gg \nu$ we estimate $S \approx -2/k_0 \ln
(k_0/\nu)$, and for $k_0 \ll \nu$ we obtain $S \approx
-k_0/(2\nu^2)$.\cite{p11}
It follows that for a given energy density there exists a seed wave number $k_{\min}$
for which the $q$-breather is most strongly localized. Moreover, $q$-breathers
tend to delocalize for $k_0 \ll k_{\min}$, that is, in the
limit of long wavelength. 
These analytical predictions may now be compared to careful numerical computations
of $q$-breathers. 

\begin{figure}
\resizebox*{0.5\columnwidth}{!}{\includegraphics{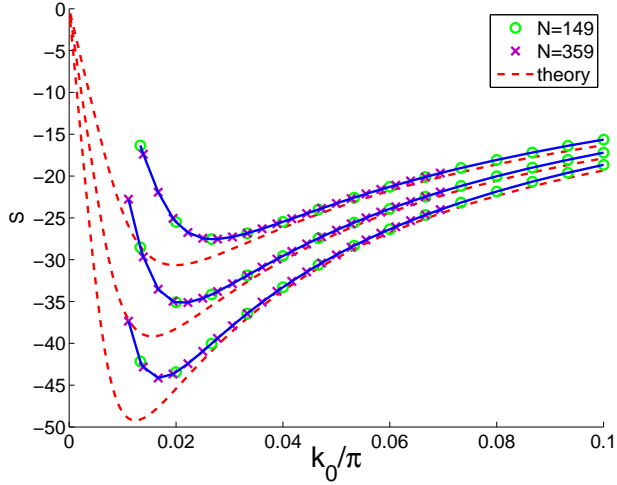}}
\caption{The slope $S(k_0)$ for $\beta=1$ and $\varepsilon = 1.57 \times 10^{-3}$, 
$9.6\times 10^{-4}$, and $6.08\times
10^{-4}$ (dashed lines from top to
bottom). The symbols and connecting lines are the results
of the numerical 
solutions of $q$-breathers for $N=149$ and
$N=359$.\cite{p11}}\label{Fig8}
\end{figure}

The analytical and numerical results are
compared in Fig.~\ref{Fig8}. 
The numerical results confirm that the 
localization length does not depend on the system size.
We also observe the
minimum $S(k_0)$ whose depth and location vary with the
energy as expected. A systematic mismatch between the theory and the numerical results 
for small $k_0$ is caused by 
higher order corrections to the perturbation theory. The $q$-breathers from
Fig.~\ref{Fig4} for $q_0=4,\, 8$ and 24 correspond to the most left
symbol on the middle curve in Fig.~\ref{Fig8}, the minimum on that curve, and a
point to the right of it, respectively.

\begin{figure}
\resizebox*{0.5\columnwidth}{!}{\includegraphics{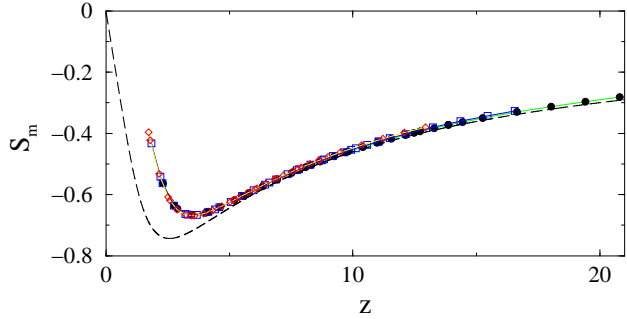}}
\caption{$S_m(z)$ (dashed line) with $z=k_0/\nu$. The symbols and connecting lines are
the rescaled data from the numerical $q$-breather solutions in
Fig.~\ref{Fig8}.\cite{p11} } \label{Fig9}
\end{figure}

It follows from Eq.~(\ref{eq13}) that the quantity $S_m(z)=\nu S$
depends on a single variable, namely the dimensionless seed wave
number $z=k_0/\nu$. 
It implies that knowing this single master slope function
is sufficient to predict the localization property of any $q$-breather,
at any energy, any seed wave number, etc.
In Fig.~\ref{Fig9} the dependence of $S_m(z)$ is
shown together with the rescaled numerical data from Fig.~\ref{Fig8}.

Despite the systematic mismatch between the
numerical data and the theoretical prediction for small $z=k_0/\pi$,
the scaled numerical data collapse to a single curve.
Therefore, the theoretically predicted scaling law in Eq.~(\ref{eq13}) is well
confirmed. 
Analogous results have been obtained for the $\alpha$-FPU model, and
for both FPU models for seed
wave numbers with frequencies near the upper edge of the
spectrum $\omega_q$.

\section{Discussion}
\label{sec6}

The reason for the delocalization of $q$-breathers at the
edges of the linear spectrum are resonances between 
normal mode frequencies: $\omega_{q_0} \approx n \omega_{nq_0}$
for low frequency modes and $\omega_{q_0} \approx \omega_{q}$ 
for high frequency modes. Thus it is expected that
the excited mode with a wave number close
to an edge of the spectrum will decay into many other modes, and
the two time scales 
$\tau_2 \approx \tau_1$. 
These time scales must diverge in the limit of small 
frequencies, due to conservation of total mechanical
momentum. Hence the time scale $\tau_{\rm sim}$ of the simulation is important. 
Suppose we excite a
normal mode that is close to a well-localized $q$-breather when $\tau_2 \gg \tau_1$.
If $\tau_{\rm sim} < \tau_2$, the excitation of this normal mode will quickly
spread into a packet, but stay
localized in normal mode space for all observation times (this localization is
the FPU problem). If $\tau_{\rm sim} > \tau_2$, the long time of nonequipartition
will be eventually replaced by a thermalized state.

Suppose we excite a low frequency normal mode for which the corresponding
$q$-breather is delocalized, and $\tau_2 \approx \tau_1$. If $\tau_{\rm sim} < \tau_2$,
the normal mode will not decay at all during the observation time.
The simulation will thus recover the linear dynamics of the bare modes
as if the nonlinearity were absent.
If $\tau_{\rm sim} > \tau_2$, the normal mode will begin to decay into other modes,
but there will be no intermediate state in which only a few other modes become excited;
rather the entire available mode space will be excited at once. That means
that these low frequency modes will behave like linear modes for short times and fully
chaotically for larger times, without any intermediate regime of 
exciting only a few other modes. 

The normal mode frequencies $\omega_q$ are periodic in the wave number $k$.
That period defines 
the irreducible part 
of the wave number space, which therefore has finite width
(for the chain
it is $\pi$). Localization in wave number
space is meaningful only when the localization length $\xi < \pi$. 
Thus, from our previous results
we conclude that all $q$-breathers which are closer than some critical
distance $\Delta k_0$ of the seed wave number $k_0$ from the
spectrum edge delocalize. For non-zero $\nu^2 \sim \beta \varepsilon$ 
$\Delta k_0\neq 0$. Thus, there exists a range
of $k_0$ for which the excitation of the normal mode decays
and the normal mode picture may become ill-defined. The
range $\Delta k_0$ is an increasing function of $\nu$, and for some critical value
$\nu$ $\Delta k_0 \approx \pi$, which means that 
normal modes do not characterize the dynamics of the system. 
That corresponds to the regime of strong interaction between
modes.
For small $\nu$ this strong interaction occurs near the edges of the
spectrum only.

Let us discuss the results for finite systems.
A given $\nu$ corresponds to a finite $\Delta k_0$. A finite
system implies a discrete set of wave numbers for the normal modes.
The thermal conductivity of anharmonic acoustic lattices is governed by the
dynamics of low frequency waves. 
To numerically observe a
conductivity different from the a free propagating (ballistic) one, 
we have to resolve the regime of strong interactions. 
In
particular, we have to choose a system size greater than $N_c \sim
1/(\Delta k_0)$. The theory we have presented allows us to obtain relevant
quantitative estimates, and the computational results show that these
considerations are valid.

Finally, let us discuss the resonance peaks in the tails
of the distributions in Fig.~\ref{Fig2}, which are observed
in the FPU trajectory and also in the $q$-breather.
A quantitative explanation of the
origin and position of these peaks was given recently.\cite{p4} They are generated by
the closeness to the resonance in Eq.~(\ref{eq7}) for certain mode
numbers. It is easy to check that the peaks appear in the upper
half of the frequency spectrum and for
low-frequency seed modes only. To explain the appearance of these peaks we have to consider
the dispersion relation more accurately.\cite{p4} Numerical experiments
show that these resonant peaks trigger the transition to
equipartition in the FPU trajectory.\cite{p4} The nature of the pathway to equipartition
for high-frequency seed modes, for which these peaks are absent,
remains a puzzle. 

There are many other questions that still wait for answers. We have to explore other nonlinear
models and test which of the results we have discussed will be generic and which not.
We need to obtain reliable estimates on the dependence of $\tau_{1,2}$ on the parameters
and compare with the results for $q$-breathers. We have to determine whether
$q$-breather-like excitations are spontaneously generated at thermal equilibrium
and to characterize their statistical properties. This list could be continued,
but there must be an end to everything, and our story finishes here. 

\begin{acknowledgments}
We thank D. Bambusi, A. Lichtenberg, L. Galgani, T.\ Penati and A. Ponno
for useful discussions.
M.I., O.K. and K.M. acknowledge support from RFBR, grant
No.~07-02-01404.
\end{acknowledgments}


\begin{thebibliography}{99}

\bibitem{via89}
V. I. Arnold, 
{\sl Mathematical Methods of Classical Mechanics}
(Springer-Verlag, New York, 1989).

\bibitem{p1} D. K. Campbell, S. Flach, and Yu. S. Kivshar, 
``Localizing energy through nonlinearity and discreteness,''
Phys. Today {\bf 57} (1)
43--49 (2004).

\bibitem{fw98pr}
S. Flach and C. R. Willis,
``Discrete breathers,''
Phys. Rep. {\bf 295}, 182--264 (1998). 

\bibitem{fg05chaos}
S. Flach and A. Gorbach,
``Discrete breathers in Fermi-Pasta-Ulam lattices,''
Chaos {\bf 15}, 015112-1--10 (2005).

\bibitem{p2}
E. Fermi, J. Pasta, and S. Ulam, 
Los Alamos Report LA-1940 (1955). Also published in 
{\sl Collected Papers of Enrico Fermi}, edited by E Segr\'e (University of 
Chicago Press, Chicago, 1965).

\bibitem{Sh} D. L. Shepelyansky, 
``Low-energy chaos in the Fermi-Pasta-Ulam problem,''
Nonlinearity {\bf 10}, 1331--1338 (1997).

\bibitem{p5} L. Berchialla, A. Giorgilli, and S. Paleari, 
``Exponentially long times to equipartition in the thermodynamic
limit,''
Phys. Lett. A
{\bf 321}, 167--172 (2004).

\bibitem{gs72}
L. Galgani and A. Scotti, 
``Planck-like distributions in classical nonlinear mechanics,''
Phys. Rev. Lett. {\bf 72}, 1173--1176 (1972).

\bibitem{njzmdk65}
N. J. Zabusky and M. D. Kruskal, 
``Interaction of solitons in a collisionless plasma and recurrence 
of initial states,''
Phys. Rev. Lett. {\bf 15}, 240--243 (1965).

\bibitem{IC} F. M. Izrailev and B. V. Chirikov, 
``Statistical properties of a non-linear chain,''
Sov. Phys. Dokl. {\bf 11}, 30 (1966).

\bibitem{Ford92}
J. Ford, 
``The Fermi-Pasta-Ulam problem: Paradox turns discovery,''
Phys. Rep. {\bf 213}, 271--310 (1992).

\bibitem{p4} T. Penati and S. Flach, 
``Tail resonances of FPU q-breathers and their impact on the pathway to equipartition,''
Chaos {\bf 17},
023102-1--16 (2007).

\bibitem{p3} S. Flach, M. V. Ivanchenko, and O. I. Kanakov, 
``q-breathers in Fermi-Pasta-Ulam chains: Existence, localization and stability,''
Phys. Rev. E {\bf 73}, 036618-1--14 (2006).

\bibitem{p6}
S. Flach, M. V. Ivanchenko, and O. I. Kanakov, 
``q-breathers and the Fermi-Pasta-Ulam problem,''
Phys. Rev. Lett. {\bf 95}, 064102-1--4
(2005).

\bibitem{p7} J. H. Conway and A. J. Jones, 
``Trigonometric diophantine equations (on vanishing sums of roots
of unity),''
Acta Arith. {\bf
XXX}, 229--240 (1976).

\bibitem{p8} M. A. Lyapunov, {\sl The General Problem of Stability of
Motion} (Taylor \& Francis, London, 1992).

\bibitem{sf04} S. Flach, ``Computational studies of discrete breathers,''
in {\sl Energy Localization and Transfer}, 
edited by T. Dauxois, A. Litvak-Hinenzon, R. S. MacKay and A. Spanoudaki (World Scientific, Singapore, 2004), pp. 1--71.

\bibitem{p9}
J. De Luca, A. J. Lichtenberg and M. A. Lieberman, 
``Time scale to ergodicity in the Fermi-Pasta-Ulam problem,''
Chaos {\bf 5}, 283--297
(1995).

\bibitem{ll92}
A. J. Lichtenberg and M. A. Liebermann, {\sl Regular and Chaotic Dynamics} (Springer, Berlin, 1992).

\bibitem{p10} M. V. Ivanchenko, O. I. Kanakov, K. G. Mishagin, and S. Flach, 
``q-breathers in finite two- and three-dimensional nonlinear acoustic lattices,''
Phys. Rev. Lett. {\bf 97}, 025505-1--4 (2006).

\bibitem{p11} O. I. Kanakov, S. Flach, M. V. Ivanchenko, and
K. G. Mishagin, 
``Scaling properties of $q$-breathers in nonlinear acoustic lattices,''
Phys. Lett. A {\bf 365}, 416--420 (2007).

\end{thebibliography}
\end{document}